\documentclass{article}
\usepackage{spconf}
\usepackage[ruled,vlined,linesnumbered]{algorithm2e}
\usepackage{amsmath, bbm}
\usepackage{amssymb}
\usepackage{amsfonts}
\usepackage{amsthm}
\usepackage{amsbsy}
\usepackage{nicefrac}
\usepackage{comment}
\usepackage{bm}
\usepackage{hyperref}
\usepackage{cite}
\usepackage[utf8]{inputenc} 
\usepackage[T1]{fontenc}
\usepackage{graphicx}
\usepackage[font=small]{caption}
\usepackage{tikz}
\usetikzlibrary{positioning}
\usepackage{pgfplots}
\usepackage{epstopdf}
\usepackage{subcaption}
\usepackage{multirow}

\def\cast{{
   \mathord{
      \hbox to 0em{
         \ooalign{
	   \smash{\hbox{$\ast$}}\crcr
	   \smash{\hskip-1pt\Large\hbox{$\circ$}} }
	 \hidewidth}
      \phantom{\bigcirc}
} }}

\def\bm#1{\mbox{\boldmath $#1$}}

\newcommand{\bds}{\begin {itemize}}
\newcommand{\eds}{\end {itemize}}
\newcommand{\bdf}{\begin{definition}}
\newcommand{\blm}{\begin{lemma}}
\newcommand{\edf}{\end{definition}}
\newcommand{\elm}{\end{lemma}}
\newcommand{\bthm}{\begin{theorem}}
\newcommand{\ethm}{\end{theorem}}
\newcommand{\bprp}{\begin{prop}}
\newcommand{\eprp}{\end{prop}}
\newcommand{\bcl}{\begin{claim}}
\newcommand{\ecl}{\end{claim}}
\newcommand{\bcr}{\begin{coro}}
\newcommand{\ecr}{\end{coro}}
\newcommand{\bquest}{\begin{question}}
\newcommand{\equest}{\end{question}}

\newcommand{\larrow}{{\larrow}}

\newcommand{\argmin}{\ensuremath{\mathrm{arg}\min}}
\newcommand{\argmax}{\ensuremath{\mathrm{arg}\max}}

\newcommand{\cG}{{\ensuremath{\mathcal{G}}}}

\newcommand{\cP}{{\ensuremath{\mathcal{P}}}}

\newcommand{\cS}{{\ensuremath{\mathcal{S}}}}

\newcommand{\vx}{{\ensuremath{{\mathbf{x}}}}}

\newcommand{\vy}{{\ensuremath{{\mathbf{y}}}}}

\newcommand{\mA}{{\ensuremath{\mathbf{A}}}}

\newcommand{\mD}{{\ensuremath{\mathbf{D}}}}

\newcommand{\mG}{{\ensuremath{\mathbf{G}}}}

\newcommand{\mI}{{\ensuremath{\mathbf{I}}}}

\newcommand{\mV}{{\ensuremath{\mathbf{V}}}}

\newcommand{\mW}{{\ensuremath{\mathbf{W}}}}

\usepackage{latexsym}

\def\IC{\mathbb C}
\def\IN{\mathbb N}
\def\IZ{\mathbb Z}
\def\IR{\mathbb R}

\def\shat{^{\mathchoice{}{}%
 {\,\,\smash{\hbox{\lower4pt\hbox{$\widehat{\null}$}}}}%
 {\,\smash{\hbox{\lower3pt\hbox{$\hat{\null}$}}}}}}

\def\bSigma{{
      \ooalign{
      \smash{\hskip.4pt\raise.4pt\hbox{$\Sigma$}}\vphantom{}\crcr
      \smash{\hskip.7pt\raise.6pt\hbox{$\Sigma$}}\vphantom{}\crcr
      \smash{\hbox{$\Sigma$}}\vphantom{$\Sigma$}}
      \vphantom{\hbox{$\Sigma$}}
      }}
\def\bTheta{{
      \ooalign{
      \smash{\hskip.5pt\raise.5pt\hbox{$\Theta$}}\vphantom{}\crcr
      \smash{\hskip.0pt\raise.1pt\hbox{$\Theta$}}\vphantom{}\crcr
      \smash{\hbox{$\Theta$}}\vphantom{$\Theta$}}
      \vphantom{\hbox{$\Theta$}}
      }}
\def\bDelta{{
      \ooalign{
      \smash{\hskip.4pt\raise.4pt\hbox{$\Delta$}}\vphantom{}\crcr
      \smash{\hskip.7pt\raise.6pt\hbox{$\Delta$}}\vphantom{}\crcr
      \smash{\hbox{$\Delta$}}\vphantom{$\Delta$}}
      \vphantom{\hbox{$\Delta$}}
      }}
\def\bLambda{{
      \ooalign{
      \smash{\hskip.5pt\raise.5pt\hbox{$\Lambda$}}\vphantom{}\crcr
      \smash{\hskip.0pt\raise.1pt\hbox{$\Lambda$}}\vphantom{}\crcr
      \smash{\hbox{$\Lambda$}}\vphantom{$\Lambda$}}
      \vphantom{\hbox{$\Lambda$}}
      }}

\makeatletter

\def\bordermatrix#1{\begingroup \m@th
  \@tempdima 8.75\p@
  \setbox\z@\vbox{%
    \def\cr{\crcr\noalign{\kern2\p@\global\let\cr\endline}}%
    \ialign{$##$\hfil\kern2\p@\kern\@tempdima&\thinspace\hfil$##$\hfil
      &&\quad\hfil$##$\hfil\crcr
      \omit\strut\hfil\crcr\noalign{\kern-\baselineskip}%
      #1\crcr\omit\strut\cr}}%
  \setbox\tw@\vbox{\unvcopy\z@\global\setbox\@ne\lastbox}%
  \setbox\tw@\hbox{\unhbox\@ne\unskip\global\setbox\@ne\lastbox}%
  \setbox\tw@\hbox{$\kern\wd\@ne\kern-\@tempdima\left[\kern-\wd\@ne
    \global\setbox\@ne\vbox{\box\@ne\kern2\p@}%
    \vcenter{\kern-\ht\@ne\unvbox\z@\kern-\baselineskip}\,\right]$}%
  \null\;\vbox{\kern\ht\@ne\box\tw@}\endgroup}
\makeatother

\makeatletter
\def\argmin{\mathop{\operator@font arg\,min}}
\def\argmax{\mathop{\operator@font arg\,max}}
\makeatother

\def\bm#1{\mbox{\boldmath $#1$}}

\newcommand{\bbeta}{{\bm\beta}}

\newcommand{\bea}{\begin{array}}
\newcommand{\ena}{\end{array}}
\newcommand{\beq}{\begin{equation}}
\newcommand{\enq}{\end{equation}}

\newcommand{\beqa}{\begin{eqnarray}}
\newcommand{\enqa}{\end{eqnarray}}

\newcommand{\beqan}{\begin{eqnarray*}}
\newcommand{\enqan}{\end{eqnarray*}}

\newcommand{\AL}{\begin{enumerate}}
\newcommand{\ALE}{\end{enumerate}}

\def\addots{\mathinner{
    \mkern1mu\raise0pt\vbox{\kern7pt\hbox{.}}
    \mkern2mu\raise4pt\hbox{.}
    \mkern2mu\raise7pt\hbox{.}
    \mkern1mu}}

\def\sddots{\mathinner{
    \mkern.8mu\raise7pt\hbox{.}
    \mkern.8mu\raise4pt\hbox{.}
    \mkern.8mu\raise0pt\vbox{\kern7pt\hbox{.}}
    \mkern1mu}}

\def\saddots{\mathinner{
    \mkern.2mu\raise0pt\vbox{\kern7pt\hbox{.}}
    \mkern.2mu\raise4pt\hbox{.}
    \mkern.2mu\raise7pt\hbox{.}
    \mkern1mu}}

\def\sqplus{\mathbin{
	{\ooalign{\hfil\raise.3ex\hbox{\scriptsize
	+}\hfil\crcr\mathhexbox274\crcr\mathhexbox275}}
	}} 
\def\sqminus{\mathbin{
	{\ooalign{\hfil\raise.3ex\hbox{\scriptsize
	--}\hfil\crcr\mathhexbox274\crcr\mathhexbox275}}
	}}

\def\IC{{
   \mathord{
      \hbox to 0em{
	 \hskip-4pt
         \ooalign{
	   \smash{\hskip1.9pt\raise2.6pt\hbox{$\scriptscriptstyle |$}}\crcr
	   \smash{\hbox{\rm\sf C}} }
	 \hidewidth}
      \phantom{\hbox{\rm\sf C}}
} }}
\def\IN{
    {\ooalign{
   \smash{\hskip2.2pt\raise1.5pt\hbox{$\scriptscriptstyle |$}}\vphantom{}\crcr
   \hbox{\sf N}
	}}
	} 
\def\IZ{
    {\ooalign{
   \smash{\hskip1.9pt\raise0pt\hbox{$\sf Z$}}\vphantom{}\crcr
   \hbox{\sf Z}
	}}
	} 
\def\IR{
    {\ooalign{
   \smash{\hskip2.2pt\raise1.5pt\hbox{$\scriptscriptstyle |$}}\vphantom{}\crcr
   \smash{\hskip2.2pt\raise3.3pt\hbox{$\scriptscriptstyle |$}}\vphantom{}\crcr
   \hbox{\sf R}
	}}
	} 

\DeclareMathAlphabet{\mathcmb}{OT1}{cmr}{b}{n}

\def\bSigma{\ensuremath{\mathcmb{\Sigma}}}
\def\bLambda{\ensuremath{\mathcmb{\Lambda}}}

\def\bTheta{\ensuremath{\mathcmb{\Theta}}}

\newcommand{\SI}{\begin{indlist}}
\newcommand{\EI}{\end{indlist}}

\newcommand{\DL}{\begin{dashlist}}
\newcommand{\DLE}{\end{dashlist}}

\makeatletter
\def\setboxz@h{\setbox\z@\hbox}
\def\wdz@{\wd\z@}
\def\boxz@{\box\z@}
\def\underset#1#2{\binrel@{#2}%
  \binrel@@{\mathop{\kern\z@#2}\limits_{#1}}}
\def\binrel@#1{\begingroup
  \setboxz@h{\thinmuskip0mu
    \medmuskip\m@ne mu\thickmuskip\@ne mu
    \setbox\tw@\hbox{$#1\m@th$}\kern-\wd\tw@
    ${}#1{}\m@th$}%
  \edef\@tempa{\endgroup\let\noexpand\binrel@@
    \ifdim\wdz@<\z@ \mathbin
    \else\ifdim\wdz@>\z@ \mathrel
    \else \relax\fi\fi}%
  \@tempa
}
\let\binrel@@\relax%
\makeatother

\newcommand\abs[1]{\left|#1\right|}

\usepackage{epstopdf}
\AppendGraphicsExtensions{.tif}

\title{Federated Deep Unfolding for Sparse Recovery}

\name{Komal Krishna Mogilipalepu$^\star$, Sumanth Kumar Modukuri$^\star$,
\thanks{This work was supported in part by the Pratiksha Trust Fellowship and SERB grant SRG/2019/000619. $^\star$These authors contributed equally to this work.}}
\secondlinename{Amarlingam Madapu and  Sundeep Prabhakar Chepuri}
\address{Indian Institute of Science, Bangalore, India}

\begin{document}
\ninept
\maketitle
\begin{abstract}
This paper proposes a federated learning technique for deep algorithm unfolding with applications to sparse signal recovery and compressed sensing. We refer to this architecture as Fed-CS. Specifically, we unfold and learn the iterative shrinkage thresholding algorithm for sparse signal recovery without transporting to a central location, the training data distributed across many clients. We propose a layer-wise federated learning technique, in which each client uses local data to train a common model. Then we transmit only the model parameters of that layer from all the clients to the server, which aggregates these local models to arrive at a consensus model. The proposed layer-wise federated learning for sparse recovery is communication efficient and preserves data privacy. Through numerical experiments on synthetic and real datasets, we demonstrate Fed-CS's efficacy and present various trade-offs in terms of the number of participating clients and communications involved compared to a centralized approach of deep unfolding.
\end{abstract}

\section{Introduction}\label{sec:intro}
Compressed sensing (CS) is a signal processing paradigm, which allows acquisition and recovery of non-bandlimited signals by utilizing the prior knowledge of the signal \cite{CS_basic1,CS_basic2}. Specifically, CS ensures exact or almost exact recovery of sparse signals from far fewer linear measurements than that is needed for Nyquist sampling \cite{CS_basic3}, thus reducing the costs and time associated with data acquisition. Hence, CS has attracted significant attention among researchers from various fields like computational and medical imaging, group testing, communication systems, and localization and positioning, to list a few \cite{CS_review}. 

Mathematically the CS framework can be modeled using a linear system of equations as $\vy=\mA\vx$, where $\vy \in \mathbb{R}^{M}$ is the measurement vector, $\mA \in \mathbb{R}^{M\times N}$ is the sensing matrix with  $M \ll N$ (hence the name compressed sensing). The inverse problem of recovering $\vx$ from $\vy$ is ill-posed, and the compression ratio $M/N$ influences the recovery. However, we can incorporate any available prior information about $\vx \in \mathbb{R}^N$ in solving this ill-posed inverse problem. In CS, the focus is on recovering a sparse vector $\vx$ with very few nonzero entries given $\mA$ and $\vy$. 

Over the last decade, a plethora of algorithms have been proposed to recover a sparse $\vx$ from $\vy$~\cite{CS_review}. Majority of the algorithms are based on iterative optimization approaches such as orthogonal matching pursuit (\texttt{OMP}) \cite{OMP}, compressive sampling matching pursuit (\texttt{CoSaMP}) \cite{cosamp}, iterative soft thresholding algorithm (\texttt{ISTA})~\cite{ISTA}, and approximate message passing algorithm (\texttt{AMP}) \cite{Amp}, to name a few. These approaches are developed from the knowledge of the model (e.g., prior knowledge about the sensing matrix), and their performance heavily depends on the proper choice of hyperparameters. Further, the iterative optimization-based approaches typically require tens to hundreds of iterations to achieve an acceptable performance.

Different from such iterative optimization-based algorithms, in \cite{deep_cs1,deep_cs2}, learning-based or data-driven approaches were proposed for sparse recovery. These approaches learn a non-linear mapping between the input and output from the training dataset 
$\{\vx_{i},\vy_{i}\}_{i=1}^{S}$ where $S$ is the number of training examples. Often, these data-driven approaches outperform iterative optimization algorithms. However, the network architecture in these approaches is empirically determined, and it is hard to interpret the network functionality.

In \cite{lista}, Gregor and LeCun introduced a novel technique called \emph{deep unfolding} for sparse coding and sparse recovery. An algorithm that produces a fast approximations for sparse coding was proposed. They also have investigated a principled framework for expressing traditional iterative algorithms (e.g., \texttt{ISTA}) as a neural network with an architecture that is interpretable. In particular, each iteration was represented as one layer of a neural network.
A deep neural network (DNN) was formed by concatenating such layers. The trained network takes measurements as input and outputs a sparse vector, thereby mimicking the iterative algorithm. 
An additional benefit is that the prior knowledge of an iterative algorithm can be transferred during the training of the network. There have been many recent advances in developing neural network architectures based on unrolling iterative algorithms for problems such as sparse recovery \cite{lista_cp}, image deblurring \cite{image_deblurring_deep_unfold}, signal recovery from one-bit quantization \cite{onebit_quantization}, and one-bit CS \cite{deep_onebit_cs_unfolding}. 

The availability of large volumes of data makes the algorithm unrolling useful and practical. Nowadays, Internet of things (IoT) devices and mobile phones are being used for collecting large volumes of data (e.g., photos, videos, location information) from various applications such as social media apps, medical equipment, and business platforms. Since the data is privacy sensitive, the cloud-centric approaches for training a neural network may not be useful \cite{fed_IoT}, where the raw data is collected from many clients to train a model. Other cloud-centric methods involve longer propagation delays and incur unacceptable latency for applications in which on-the-edge real-time decisions have to be made. Alternatively, for training a complex collaborative machine learning (ML) model by guaranteeing that the training data remains on personal devices, a decentralized ML approach called \emph{federated learning} (FL) was introduced in \cite{Fed_basic}. An important advantage of the FL approach is the decoupling of model training from the essence of direct access to the raw training data \cite{fed_medicin,fed_wireless,fed_IoT}. 

Existing learning-based sparse recovery approaches are developed with a peer-to-peer setting, where measurements of a sparse vector are taken on different devices, which then sends it to a central unit. The decoding of the underlying sparse vector is done either centrally or locally based on the model that is trained centrally. For many devices that require on-the-edge decision making, algorithm unrolling using concepts from FL becomes imperative. Therefore, in this paper, we propose a federated algorithm unfolding approach for learning the iterations of a sparse recovery algorithm called \texttt{ISTA}. We refer to this as \texttt{Fed-CS}. Precisely, we propose an algorithm that leverages advantages from algorithm unfolding and FL. This is the first work that combines both algorithm unfolding and federated learning for solving sparse recovery problems to the best of our knowledge. 

Our contributions are two-fold: i) We propose a novel algorithm for a layer-wise federation of a layer-wise trained model that learns, ii) We demonstrate the performance of our method with extensive numerical experiments on synthetic and real datasets and compare with the state-of-the-art approaches.  
\vspace*{-2mm}
\section{preliminaries}
In this section, we give a brief description on algorithm unfolding for sparse recovery.
\vspace*{-3mm}
\subsection{Learning sparse recovery iterations}
We can recover the sparse vector $\vx$ from the compressed observations $\vy=\mA\vx$ by solving the convex program~\cite{CS_basic1}
\begin{equation}
\underset{{\vx} \in \mathbb{R}^N}{\rm minimize} \,\, {\|\vy-\mA\vx\|_{2}^{2}+\lambda\|\vx\|_{1}},
\tag{$\cP_1$}
\end{equation}
where $\|\vx\|_1 = \sum_{i=1}^N \abs{x_i}$ is the $\ell_1$ norm and $\lambda$ is a hyper parameter that controls the sparsity. The problem $(\cP_1)$ does not admit a closed-form solution and hence solved iteratively. \texttt{ISTA} \cite{ISTA} is one of the commonly used algorithms to iteratively solve the problem $(\cP_1)$. The update equation of \texttt{ISTA} is given by
\begin{equation}
 \vx^{(i+1)}=\sigma_{\lambda t} \left[\vx^{(i)}+t \mA^{T}\left(\vy-\mA\vx^{(i)}\right) \right]    
\end{equation}
where  $[\sigma_{\theta}(\vx)]_i := [\abs{x_i}-\theta]_+\operatorname{sign}(x_i)$ is the soft thresholding (shrinkage) function, and $t$ is an appropriate stepsize.

Learned ISTA (LISTA) \cite{lista} unfolds the ISTA iterations and trains each iteration as a layer of DNN by treating $\mV^{(i)}= t \mA^{T} \in \mathbb{R}^{N \times M}$, $\mW^{(i)}= \mI-t \mA^{T}\mA \in \mathbb{R}^{N \times N}$, and ${\theta}^{(i)}= \lambda t \in \mathbb{R}_+$ as the trainable network parameters. Specifically, the $i$th layer of the \texttt{LISTA} network is expressed as
 \begin{equation}
  \vx^{(i+1)}=\sigma_{\theta^{(i)}} \left[\mV^{(i)}\vy+\mW^{(i)}\vx^{(i)}\right], 
  \label{eq:lista}
 \end{equation}
where ${\boldsymbol\Phi}^{(i)} = \{\mV^{(i)},\mW^{(i)},\theta^{(i)}\}$ collects the network parameters of the $i$th layer that is learnt from the training data and $\sigma_{\theta^{(i)}}[\cdot]$ is the nonlinearity.

Let us define the trainable network parameters for a network with $L$ layers as ${\boldsymbol\Theta}^{(l)} =\{{\boldsymbol\Phi}^{(i)}\}_{i=1}^{l}$ and $\boldsymbol{\Theta} = \boldsymbol{\Theta}^{(L)}$. 
Then the \texttt{LISTA} network with $L$ layers can be interpreted as an estimator $\hat{\vx} = \cG(\vy, \vx^{(0)};{\boldsymbol\Theta}^{(L)})$, parameterized by $\boldsymbol{\Theta}^{(L)}$, that estimates the unknown sparse signal $\vx$ given $\vy$ and an initial point $\vx^{(0)}$. Given the training dataset $\{(\vx_s,\vy_s)\}_{s=1}^S$, the \texttt{LISTA} network is trained by minimizing the loss function \begin{align}
f({\boldsymbol\Theta}^{(L)}) &=   \sum_{s\in \cS} \sum_{i=1}^{L} \|\vx_s- \cG(\vy_s, \vx^{(0)}_s; {\boldsymbol\Phi}^{(i)} )\|_{2}^{2} \nonumber\\
& = \sum_{i=1}^{L} f^{(i)}({\boldsymbol\Phi}^{(i)}) =  \sum_{s\in \cS} f_s({\boldsymbol\Theta}^{(L)}),\label{eq:loss2}
\end{align}
where $\cS = \{1,2,\cdots, S\}$ and $\cG(\vy, \vx^{(0)};\boldsymbol{\Phi}^{(i)}) =: \hat{\vx}^{(i)}$ is the estimated sparse vector at $i$th layer of the network. We can train the \texttt{LISTA} network by minimizing the loss $f(\boldsymbol{\Theta}^{(L)})$ with respect to the network parameters $\boldsymbol{\Theta}^{(L)}$, e.g., using stochastic gradient descent as
$
{\boldsymbol \Theta}^{(L)} \gets {\boldsymbol \Theta}^{(L)}-\alpha \nabla f({\boldsymbol \Theta}^{(L)}).
$
Here, $\alpha$ is the learning rate.

\begin{figure}[!t]
    \centering
    \includegraphics[scale=0.3,trim=1cm 2cm 1.5cm 0cm]{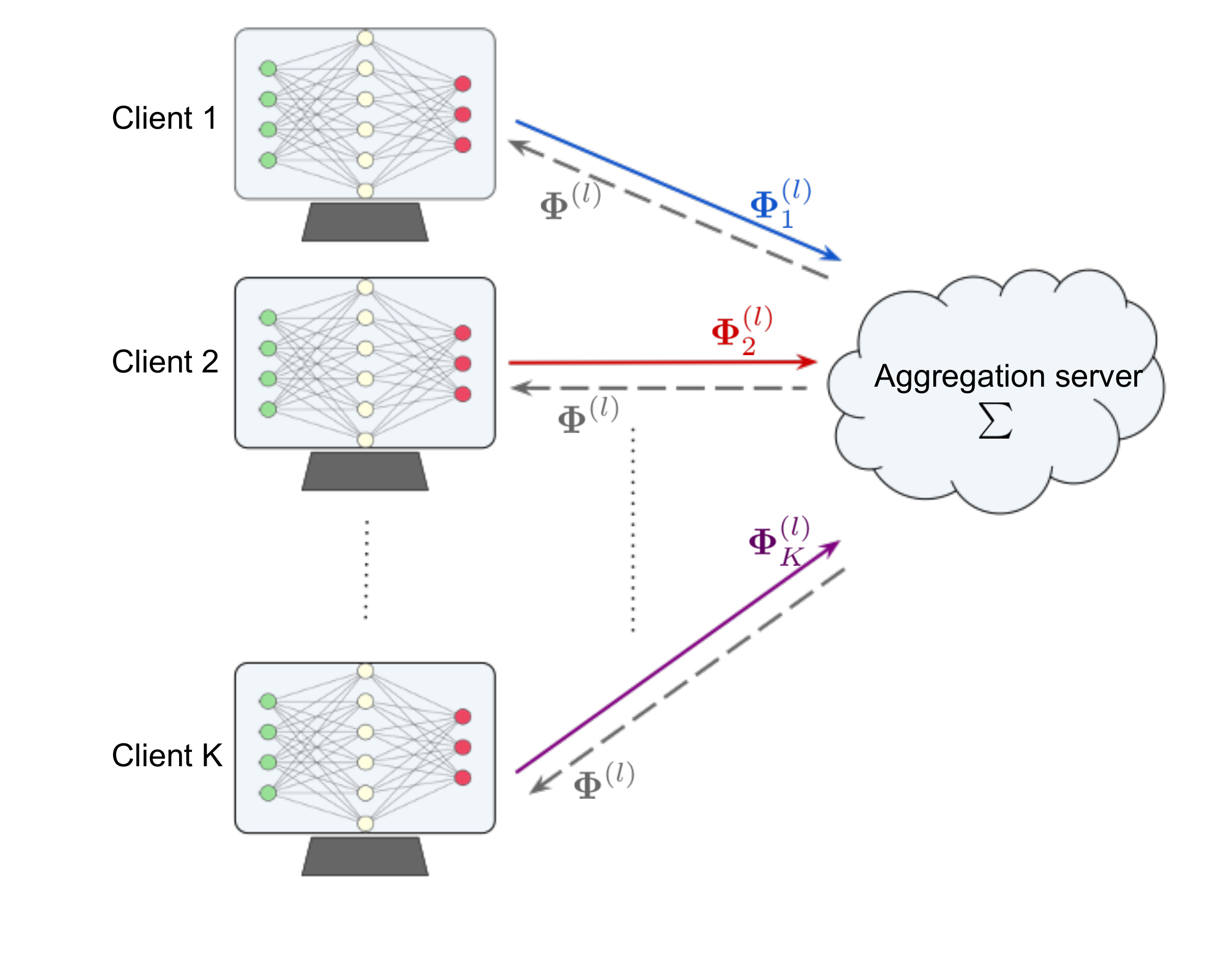}
    \caption{Layer-wise federated learning for sparse recovery.}
    \label{fig:Fed_learning}
    \vspace*{-3mm}
\end{figure}

\vspace*{-3mm}
\subsection{Layer-wise training}
 Layer-wise training is a commonly used technique for training unfolded neural networks \cite{lista_cp}.  Let us define the initial learning rate as $\alpha_{0}$ and the decayed learning rates $\alpha_{1}$ and $\alpha_{2}$. The layer-wise training is then performed in $3$ stages, and the $l$th layer is trained as follows:
 First, we train the network parameters ${\boldsymbol \Phi}^{(l)}$ by minimizing the loss function $f^{(l)}({\boldsymbol \Phi}^{(l)})$ with the initial learning rate $\alpha_{0}$. Next, use the pre-trained weights to form ${\boldsymbol \Theta}^{(l)} = {\boldsymbol \Theta}^{(l-1)} \cup {\boldsymbol \Phi}^{(l)}$, and train the network by minimizing the loss $f({\boldsymbol \Theta}^{(l)})$ with learning rates $\alpha_1$ and $\alpha_2$. Finally, we multiply each weight in ${\boldsymbol \Theta}^{(l)}$ with a decaying rate $\beta$.
 We then proceed to training the weights of the next layer (See Algorithm~\ref{alg:local_training} that we discuss later for details). The decaying learning rates ensures that the parameters of the previous layers changes slowly when the training proceeds to deeper layers~\cite{lista_cp}. 
 
Next, we describe the proposed architecture for federated algorithm unfolding for sparse recovery (\texttt{Fed-CS}), in which we depart from a centralized learning procedure to a distributed learning approach.

\vspace*{-4mm}
\section{Fed-CS architecture}

Federated learning enables a collaborative approach to train a machine learning (ML) models, where multiple collaborators, referred to as clients, train the same model in parallel on their local dataset. Further, each client sends their updated models to a central server, which then aggregates these client models into a consensus model. The server then sends back the consensus model to all clients for further training or deployment. This process is depicted in Fig.~\ref{fig:Fed_learning}. Each iteration of this process or the so-called communication round includes parallel training at clients, aggregating at servers and distributing back the model to the clients. In this work, we specialize federated learning for learning ISTA updates. Although we restrict ourselves to the case where all the clients participate synchronously, extension to the case of asynchronous client participation can be done along the similar lines. 

Consider a setup with $K$ clients and a server as shown in Fig.~\ref{fig:Fed_learning}. Let us partition the set $\cS$ into mutually disjoint non-empty sets $\cS_k$ such that $\cS = \bigcup_{k=1}^K \cS_k$ and $\cS_k \cap \cS_l = \emptyset$ for $k \neq l$. Suppose  the training data available at the $k$th client is $\{(\vx_s,\vy_s), \forall s \in \cS_k\}$. At each client, we train a local network of the form \eqref{eq:lista}, i.e.,
 \begin{equation}
  \vx^{(i+1)}_k=\sigma_{\theta^{(i)}_k} \left[\mV_k^{(i)}\vy+\mW_{k}^{(i)}\vx^{(i)}\right] 
  \label{eq:lista_client}
 \end{equation}
with trainable parameters of the $i$th layer $\mV_k^{(i)} \in \mathbb{R}^{N \times M}$, $\mW_k^{(i)} \in \mathbb{R}^{N \times N}$, and ${\theta^{(i)}_k} \in \mathbb{R}_+$. Let us collect these parameters related to the $k$th client for the $l$th layer and all the $L$ layers in $\boldsymbol{\Phi}_k^{(i)} = \{\mV_k^{(i)}, \mW_k^{(i)},{\theta^{(i)}_k}\}$ and ${\boldsymbol\Theta}^{(L)}_k =\{{\boldsymbol\Phi}^{(i)}_k\}_{i=1}^{L}$, respectively.

Since the gradient of $f(\boldsymbol{\Theta}^{(L)})$ in \eqref{eq:loss2} with respect to $\boldsymbol{\Theta}^{(L)}$ can be expressed as
\begin{align*}
\nabla f(\boldsymbol{\Theta}^{(L)}) =  \sum_{s \in \cS} \nabla f_s(\boldsymbol{\Theta}^{(L)}) &=  \sum_{k=1}^K \frac{|\cS_k|}{|\cS|} \sum_{s \in \cS_k}\nabla f_s(\boldsymbol{\Theta}^{(L)}) \\
&= \sum_{k=1}^K \sum_{l=1}^L\frac{|\cS_k|}{|\cS|} \sum_{s \in \cS_k}\nabla f_s^{(l)}(\boldsymbol{\Phi}^{(l)}).
\end{align*}
The gradients computed at the clients based on their local data, denoted by 
$
\mG_k =  \sum_{s \in \cS_k}\nabla f_s(\boldsymbol{\Theta}^{(L)})
$
can be aggregated at the server to compute the consensus gradient $\nabla f(\boldsymbol{\Theta}^{(L)})$. This means that, we can equivalently compute the update $\boldsymbol{\Theta}$ by aggregating the local updates at the clients as
\begin{eqnarray}
\text{Local training:}& \quad \quad {\boldsymbol \Theta}_{k} &\gets \quad {\boldsymbol \Theta}_k-\alpha \mG_{k} \\
\text{Aggregation:}& \quad \quad {\boldsymbol \Theta} &\gets \quad  \sum_{k=1}^{K}  \frac{|\cS_k|}{|\cS|} {\boldsymbol \Theta}_k.
\end{eqnarray}
In other words, each client computes one step of gradient descent on the current model parameters with its local data, and the server then takes a weighted average of the all the client's model parameters and sends it back to the clients.     
\begin{algorithm}[!t]
\caption{Local layer-wise training at the client}\label{alg:local_training}
  \SetAlgoLined
\texttt{training}($k$, $l$, $\boldsymbol{\Theta}^{(l-1)}_k$, $\boldsymbol{\Phi}^{(l)}$, $\alpha_0$, $\alpha_1$, $\alpha_2$, $\beta$, $E$)

$\boldsymbol{\Phi}_k^{l} \gets \boldsymbol{\Phi}^{(l)}$

\For{$e=1$ to $E$}
{
${\boldsymbol \Phi}_{k}^{(l)} \gets {\boldsymbol \Phi}_{k}^{(l)} - \alpha_0 \sum_{s \in \cS_k} \nabla f_s^{(l)}({\boldsymbol \Phi}_{k}^{(l)})$
}

$\boldsymbol{\Theta}_k^{(l)} \gets \boldsymbol{\Theta}^{(l-1)}_k \cup {\boldsymbol \Phi}_{k}^{(l)}$

\For{$e=1$ to $E$}{
${\boldsymbol \Theta}_{k}^{(l)} \gets {\boldsymbol \Theta}_{k}^{(l)} - \alpha_1 \sum_{s \in \cS_k} \nabla f_s({\boldsymbol \Theta}_{k}^{(l)})$
}

\For{$e=1$ to $E$}{
${\boldsymbol \Theta}_{k}^{(l)} \gets {\boldsymbol \Theta}_{k}^{(l)} - \alpha_2 \sum_{s \in \cS_k} \nabla f_s({\boldsymbol \Theta}_{k}^{(l)})$
}

${\boldsymbol \Theta}_{k}^{(l)} \gets \beta {\boldsymbol \Theta}_{k}^{(l)}$

Return \{${\boldsymbol \Theta}_{k}^{(l-1)}, {\boldsymbol \Phi}_{k}^{(l)}\}$
\end{algorithm}
\vspace*{-3mm}
\subsection{Layer-wise federated learning for sparse recovery}

As mentioned in the previous section, in an unfolded neural network, every layer represents an iteration, and increase in the number of layers improves the model performance, in general~\cite{lista}. More importantly, the number of model parameters that we communicate to the server is significantly less when we combine federate learning with the layer-wise training procedure. To do so, without loss of generality, we fix the number of communication rounds per layer to $C$ and the number of local epochs at each client to $E$. The three stage layer-wise training at each client is described in Algorithm~\ref{alg:local_training}.

Next, we communicate the locally trained $l$th layer parameters $\boldsymbol{\Phi}_k^{(l)}$, $k=1,2,\ldots,K$ from all the clients to the server, which then aggregate these parameters to arrive at the consensus model for the $l$th layer. To arrive at a consensus the aggregation is performed over $C$ communication rounds. After training $L$ layers in this distributed fashion, we perform one final aggregation step that aggregate $\boldsymbol{\Theta}_k^{(L)}$, $k=1,2,\ldots,K$ to arrive at the consensus model ${\boldsymbol{\Theta}}$. The entire procedure is summarized as Algorithm~2.

\begin{algorithm}[!t]

  \SetAlgoLined
  \KwIn{$K$: number of clients, $C$: number of communication rounds, $E$: number of local epochs, $\alpha_{0}, \alpha_{1},\alpha_{2}$: learning rates and $\beta$:~decaying rate.}\
      Initialize $\{{\boldsymbol \Theta}^{(0)}_1, {\boldsymbol \Theta}^{(0)}_2,\cdots,{\boldsymbol \Theta}^{(0)}_K\}$
      
   \For{$l=1$ to $L$}{
        Initialize $\boldsymbol{\Phi}^{(l)}$
        
   \For{$c =1$ to $C$}{
     \For{$k=1$ to $K$}{
         $\{ \boldsymbol{\Theta}_{k}^{(l-1)}, \boldsymbol{\Phi}_{k}^{(l)}\}$
         $\gets$ 
         $\texttt{training}$($k$, $l$, $\boldsymbol{\Theta}^{(l-1)}_k$, $\boldsymbol{\Phi}^{(l)}$, $\alpha_0$, $\alpha_1$, $\alpha_2$, $\beta$, $E$)
         
         $\boldsymbol{\Theta}_k^{(l)} \gets \boldsymbol{\Theta}^{(l-1)}_k \cup {\boldsymbol \Phi}_{k}^{(l)}$
         
         \texttt{Send} $\boldsymbol{\Phi}_{k}^{(l)}$ to server
         
         }
         $\boldsymbol{\Phi}^{(l)} \gets \sum_{k=1}^{K} \frac{|\cS_{k}|}{|\cS|} \boldsymbol{\Phi}_{k}^{(l)}$} 
            }
    \For{$k=1$ to $K$}{
 
         \texttt{Send} $\boldsymbol{\Theta}_{k}^{(L)}$ to the  server
         
         }        
  $\boldsymbol{\Theta} \gets \sum_{k=1}^{K} \frac{|\cS_{k}|}{|\cS|} \boldsymbol{\Theta}_{k}^{(L)}$         

 \caption{Fed-CS: layer-wise training}
\end{algorithm}

\begin{figure*}[h]
	\centering
        \begin{subfigure}[b]{0.25\textwidth}
        
                \includegraphics[width=\linewidth]{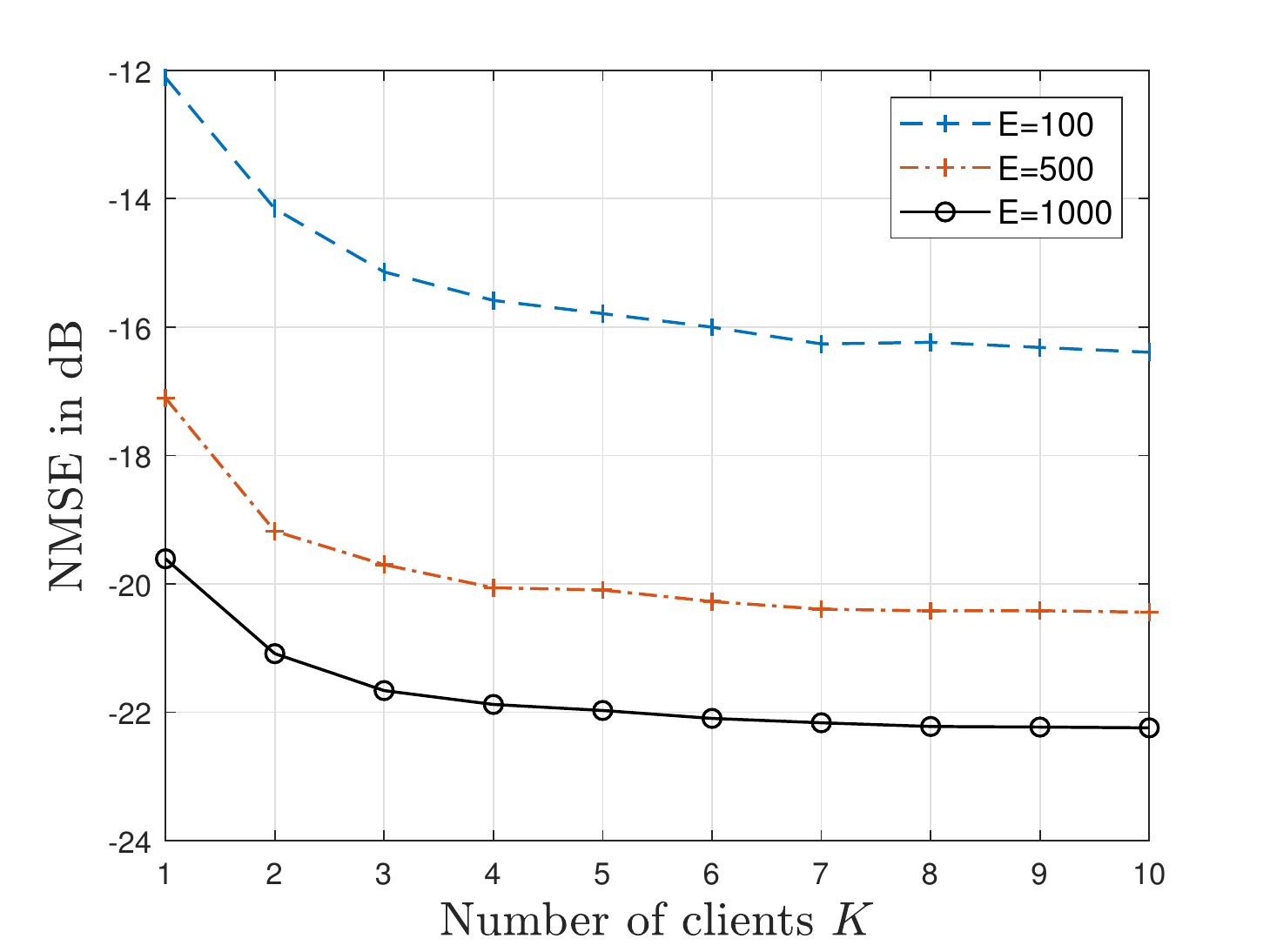}
                \caption{}
               \label{fig:clients}
        \end{subfigure}%
        \begin{subfigure}[b]{0.25\textwidth}
        
                \includegraphics[width=\linewidth]{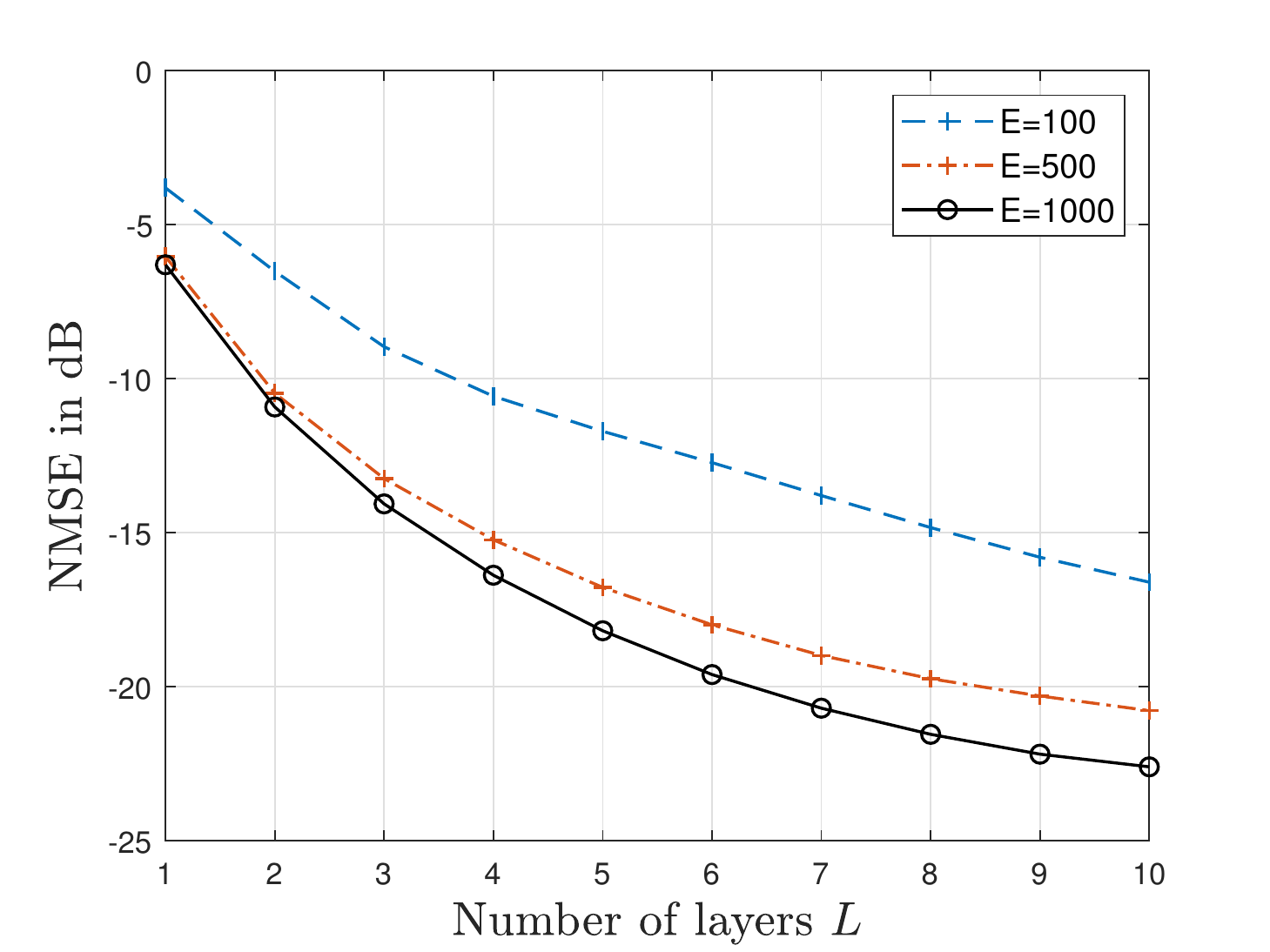}
                \caption{}
                \label{fig:layer}
        \end{subfigure}%
        \begin{subfigure}[b]{0.25\textwidth}
        
                \includegraphics[width=\linewidth]{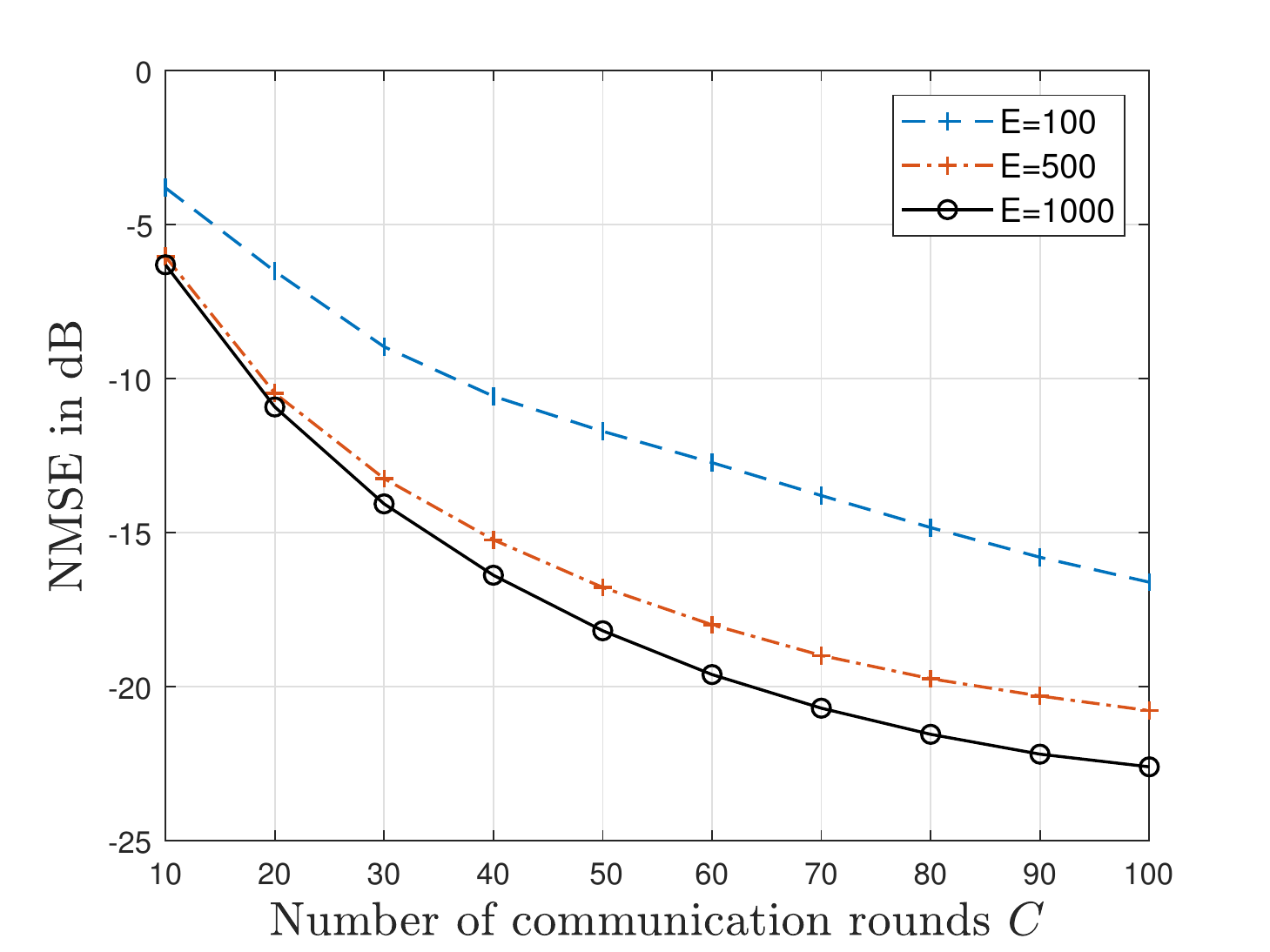}
                \caption{}
                \label{fig:com-rounds}
        \end{subfigure}%
        \begin{subfigure}[b]{0.25\textwidth}
        
                \includegraphics[width=\linewidth]{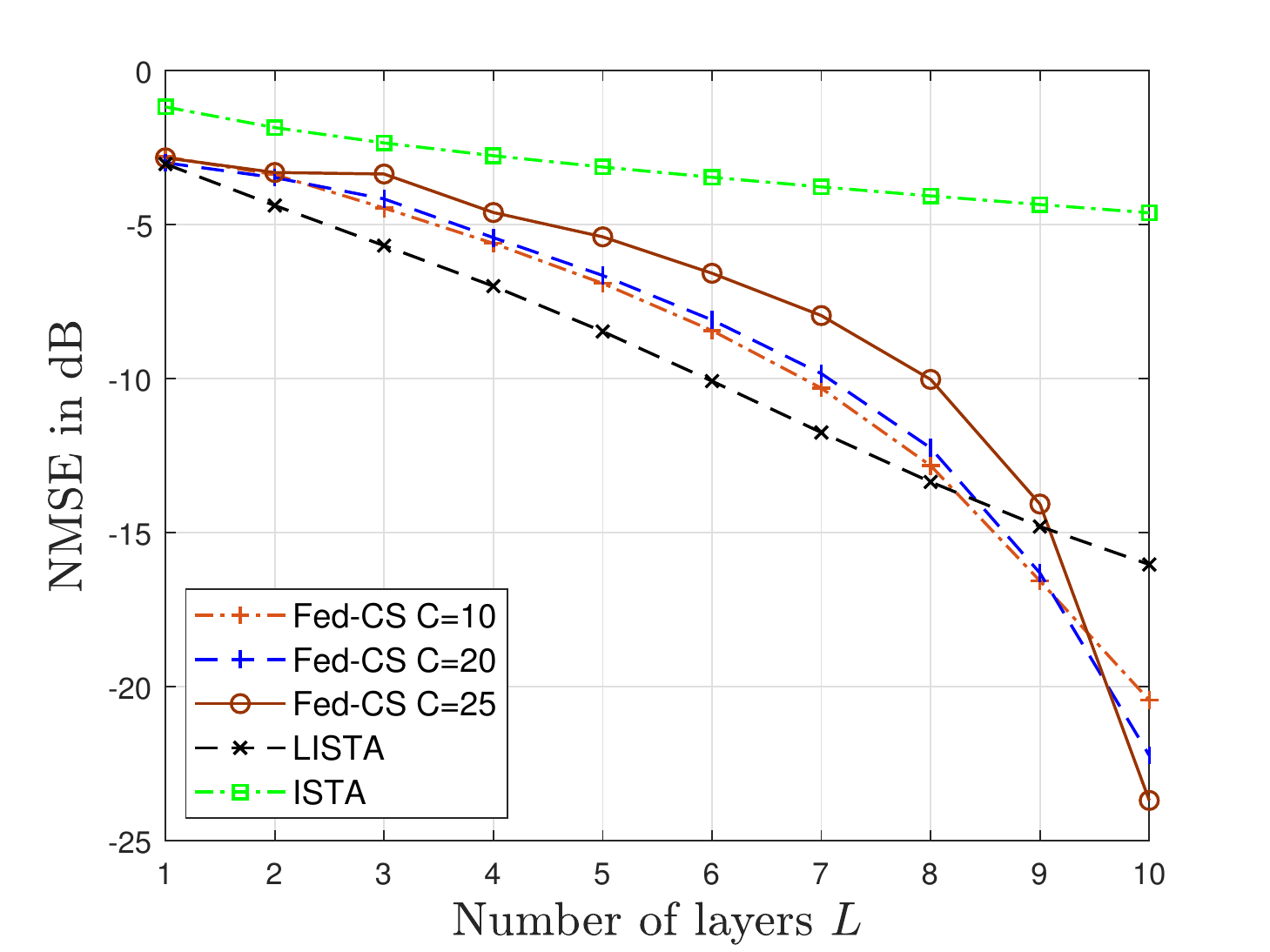}
                \caption{}
                \label{fig:comparision}
        \end{subfigure}%
\caption{Evaluation of $\texttt{Fed-CS}$ on synthetic data with respect to different number of clients $K$, layers $L$, communication rounds $C$, and local epochs~$E$.}
\vspace*{-6mm}
\end{figure*}

\begin{figure*}[h]
     \centering
      \begin{subfigure}[b]{0.3\columnwidth}
         \centering
         \includegraphics[width=\linewidth]{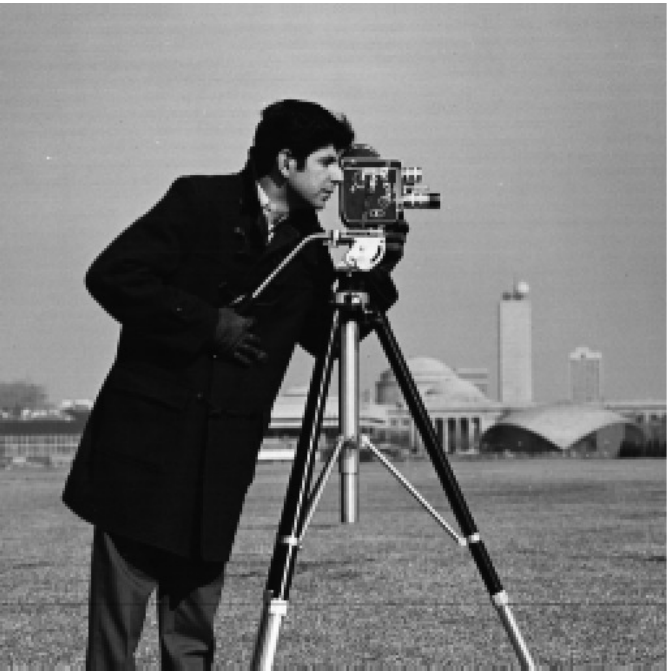}
         \caption{Ground truth}
         \label{fig:cameraman_ista}
     \end{subfigure}
    ~
     \centering
     \begin{subfigure}[b]{0.3\columnwidth}
         \centering
         \includegraphics[width=\linewidth]{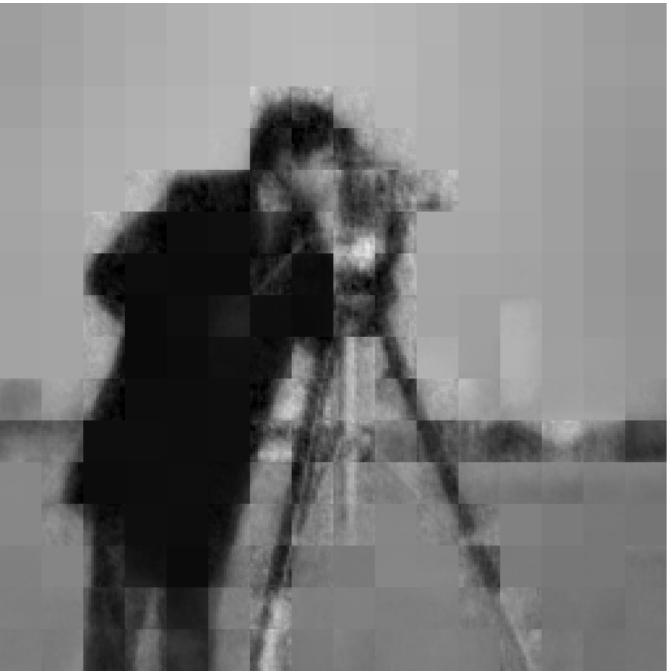}
         \caption{ISTA}
         \label{fig:cameraman_ista}
     \end{subfigure}
     ~
     \begin{subfigure}[b]{0.3\columnwidth}
         \centering
         \includegraphics[width=\linewidth]{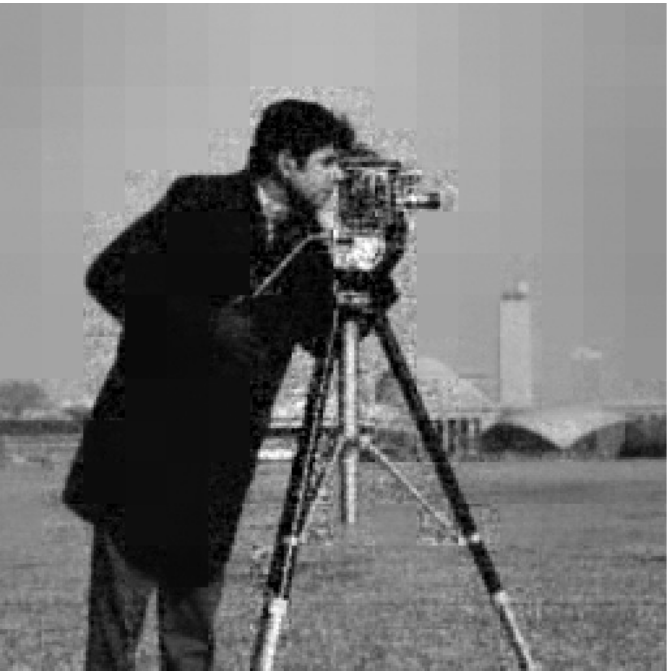}
         \caption{LISTA}
         \label{fig:cameraman_lista}
     \end{subfigure}
     ~
     \begin{subfigure}[b]{0.3\columnwidth}
         \centering
         \includegraphics[width=\linewidth]{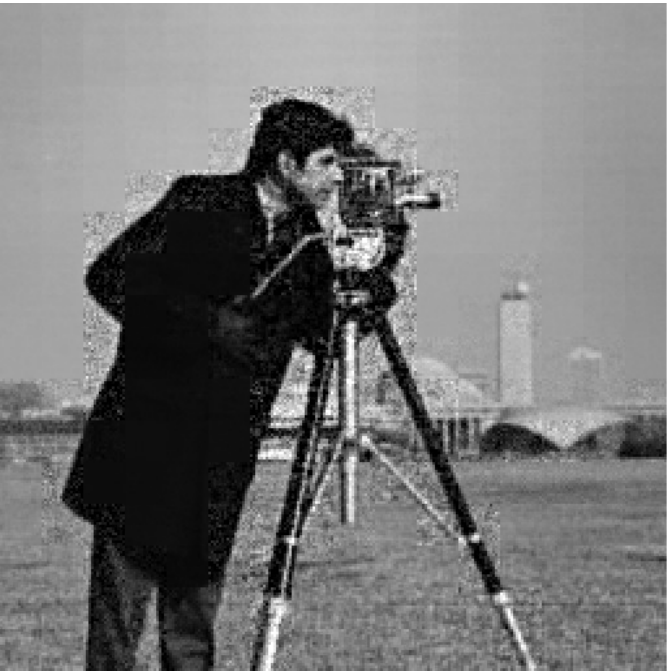}
         \caption{Fed-CS}
         \label{fig:cameraman_fed_lista}
     \end{subfigure}\\     
      \begin{subfigure}[b]{0.3\columnwidth}
         \centering
         \includegraphics[width=\linewidth]{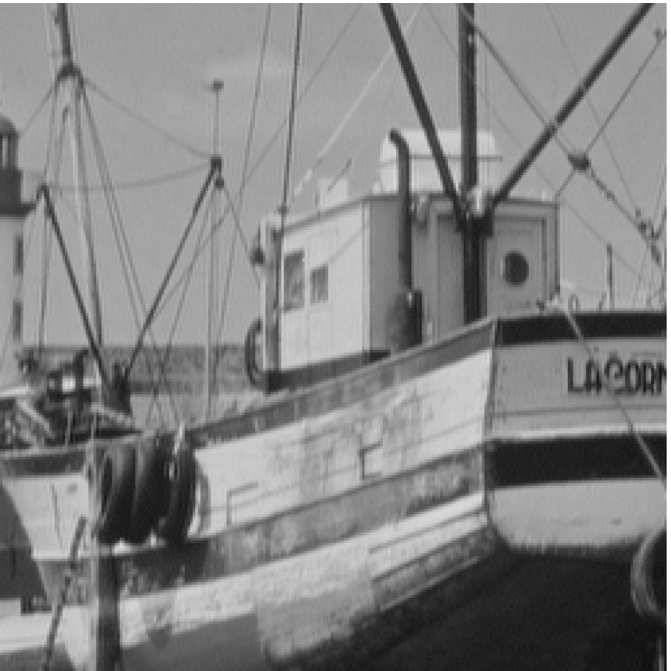}
         \caption{Ground truth}
         \label{fig:cameraman_ista}
     \end{subfigure}
    ~
     \centering
     \begin{subfigure}[b]{0.3\columnwidth}
         \centering
         \includegraphics[width=\linewidth]{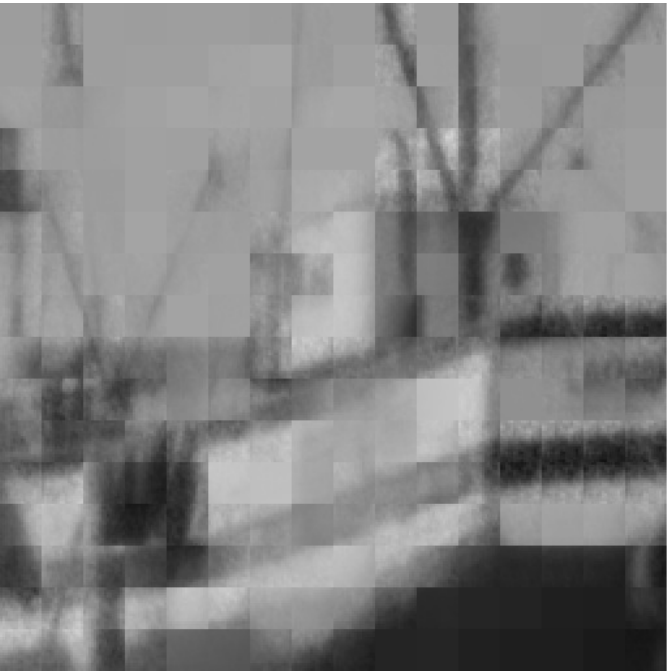}
         \caption{ISTA}
         \label{fig:cameraman_ista}
     \end{subfigure}
     ~
     \begin{subfigure}[b]{0.3\columnwidth}
         \centering
         \includegraphics[width=\linewidth]{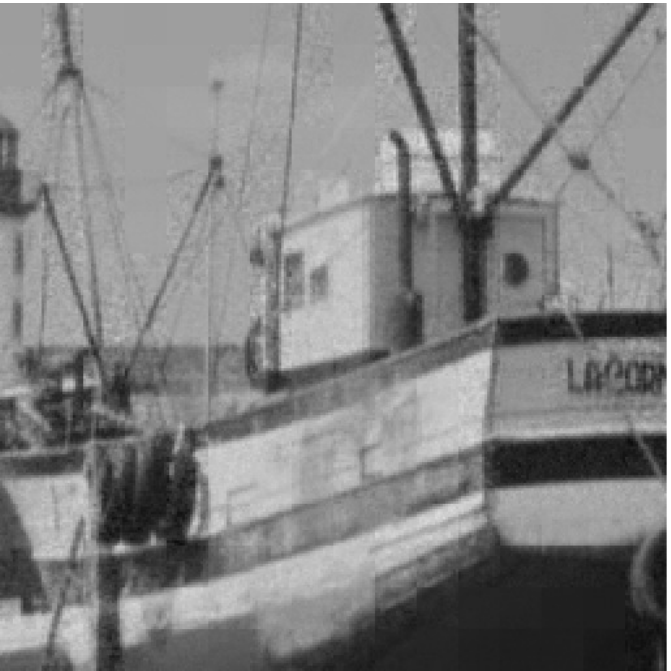}
         \caption{LISTA}
         \label{fig:cameraman_lista}
     \end{subfigure}
     ~
     \begin{subfigure}[b]{0.3\columnwidth}
         \centering
         \includegraphics[width=\linewidth]{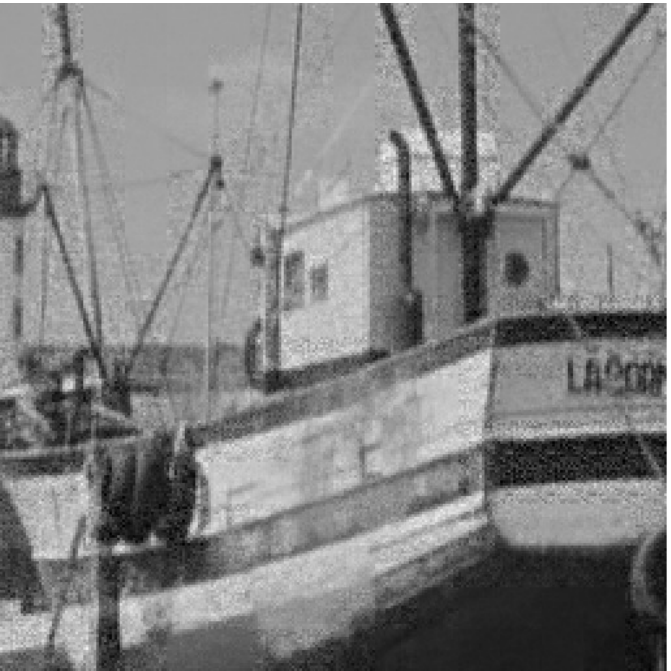}
         \caption{Fed-CS}
         \label{fig:cameraman_fed_lista}
     \end{subfigure}     
\caption{Reconstructed sample images using \texttt{ISTA}, \texttt{LISTA} and \texttt{FED-CS} approaches with 40\% compression rate. }        \label{fig:test_image}
\vspace*{-6mm}
\end{figure*}

\vspace*{-2mm}
\section{Numerical experiments}
In this section, we present results from numerical experiments on synthetic and real datasets. 

\textbf{Synthetic dataset:} We consider a sensing matrix $\mA$ with $M$ = $250$ and $N$ = $500$. The elements of $\mA$ are i.i.d. and generated using a Gaussian distribution, i.e., $a_{ij}\sim\mathcal{N} (0,\frac{1}{M})$. The columns of the sensing matrix $\mA$ are normalized to have a unit $\ell_{2}$ norm. We use this $\mA$ for all the experiments. To generate a sparse  vectors ${\vx}$, we choose its entry to be non-zero according to the Bernoulli distribution with $p=0.1$ (i.e., on average we generate sparse vectors with $10$\% non-zero entries). A test set of $1000$ samples is generated and fixed for all the experiments. We consider the number of clients $K = 10$, communications $C = 10$, and evaluate the model for $L = 10$ layers. We use the tuned learning rates $\alpha_{0}=5e^{-4}$, $\alpha_{1}=0.2\alpha_{0}$, $\alpha_{2}=0.02\alpha_{0}$, and the decaying rate $\beta=0.3$. We consider normalized mean squared error (NMSE) in dB=$10 \log_{10}$( $\mathbb{E}\|\vx-\hat{\vx}\|_{2}^{2}/\mathbb{E}\|\vx\|_{2}^{2}$) as the performance metric for sparse recovery. For a comparative analysis, \texttt{LISTA}~\cite{lista} and \texttt{ISTA}~\cite{ISTA} are considered. For a fair comparison, we set the number of iterations for \texttt{ISTA} equals to $L$, as in \texttt{LISTA} and \texttt{Fed-CS}. 

For evaluating the performance with respect to the number of clients, we train $\texttt{Fed-CS}$ by varying $K$ from $1$ to $10$, where each client has fixed number $|\cS_{k}|$ of examples. We train \texttt{Fed-CS} for different number of local epochs $E={100, 500, 1000}$.  Fig.~\ref{fig:clients}, shows the NMSE, averaged over the considered $1000$ test data points for different number of clients. We see that the performance of \texttt{Fed-CS} improves with the number of clients. Further, from Fig.~\ref{fig:clients}, the results indicates that the performance of the proposed \texttt{Fed-CS} improves as number of local epochs $E$ per client increases.     



        
        

%

Next, we present the performance of the model with respect to the number of layers. Note that for evaluating the performance of the network we compute the NMSE of the model after each layer is trained. In \texttt{Fed-CS}, the number of communication rounds are fixed per a layer, hence the performance can be visualized in terms of the communication rounds. Fig.~\ref{fig:layer} and Fig.~\ref{fig:com-rounds} (which is same as Fig.~\ref{fig:layer} but with different x-axis) show the performance of the \texttt{Fed-CS} improves with number of layers or communication rounds, respectively. 
Finally, we compare $\texttt{Fed-CS}$ with the traditional approaches, namely, \texttt{ISTA} and $\texttt{LISTA}$. Here, we set $E=500$ and the communication rounds $C=\{10, 20, 25\}$. Fig.~\ref{fig:comparision} shows the NMSE for different layers, where we show the layer-wise performance of the trained model for $L=10$ layers. From Fig.~\ref{fig:comparision}, we can observe that the proposed \texttt{Fed-CS} can outperform the traditional \texttt{ISTA}~\cite{ISTA} and \texttt{LISTA}~\cite{lista} as the number of layers increases. Furthermore, from Fig.~\ref{fig:comparision}, we can see the performance improves with the increase in the communication rounds.

\textbf{Real dataset:} We consider the natural image dataset BSD500\cite{BSD500}, which contains $500$ natural images. We use $400$ images for training, $50$ images for validation, and $50$ images for testing. As natural images are not sparse in the spatial domain, we construct a dictionary $\mD \in \mathbb{R}^{256\times512}$ using the block proximal gradient algorithm~\cite{block_proximal}. For training the dictionary $\mD$, we extract $10000$ patches of size of $16\times16$ from the $400$ training images. We consider $\mA=\boldsymbol{\Psi}\mD$, where $\boldsymbol{\Psi} \in \mathbb{R}^{M \times 256}$ is now the sensing matrix, which is generated as described before for the synthetic dataset. We consider $E=500$, $C=10$, $K=5$, and test the trained \texttt{Fed-CS} for $L=10$ layers. 

To evaluate \texttt{Fed-CS} on the real dataset, we consider about $40$\% compression with $M=102$ for training and testing. Fig.~\ref{fig:test_image}, illustrates the recovered cameraman and boat images using the proposed \texttt{Fed-CS} and baseline approaches. We can see that the recovery of the proposed approach is comparable to \texttt{LISTA} and outperforms \texttt{ISTA}, where \texttt{LISTA} and \texttt{ISTA} use a centralized training technique. We compute PSNR by averaging over $11$ test images. The averaged PSNR values for \texttt{ISTA}, \texttt{LISTA}, and \texttt{Fed-CS} are $22.45$, $29.05$ and $27.44$, respectively. In terms of the average PSNR, the performance of \texttt{Fed-CS} is again comparable to \texttt{LISTA}. 

Before concluding this section, we emphasize that \texttt{Fed-CS} is not a competing technique to \texttt{LISTA}, but using more local epochs, clients, or communication rounds the consensus model may benefit from distributed data.

\vspace*{-5mm}
\section{Conclusions}
We presented a layer-wise federated learning technique for sparse signal recovery via deep unfolding. The proposed method is useful in scenarios with many training examples that are distributed in different locations. Unlike the centralized deep unfolding methods, the proposed federated deep unfolding method does not require transporting all the training data to a central location. We also have demonstrated using real and synthetic datasets that the consensus model obtained by aggregating the locally trained models performs on par with the centrally trained models for sparse recovery.

\pagebreak
\bibliographystyle{IEEEbib}
\bibliography{Fed_CS}

\end{document}